\documentclass[12pt]{preprint}
\let\mycapt=\caption
\let\phi=\varphi
\let\rho=\varrho
\def\caption#1{\parbox{\hsize}{\mycapt{\small{#1}}}}
\usepackage[dvips]{graphicx}
\usepackage{amsfonts}
\usepackage{amssymb}\usepackage{times}
\usepackage{sub_JP}
\usepackage{MHequ}

\newcommand{\rate}[2]{j_{#1 | #2}}
\def\A{{\cal A}}
\def\OO{{\cal O}}

\def\rate{J}
\def\L{{\rm L}}
\def\R{{\rm R}}
\def\J{{\rm J}}
\def\j{{\rm j}}
\def\Q{{\rm Q}}
\def\G{{\rm G}}

\let\realrho=r
\let\epsilon=\varepsilon
\let\kappa=\varkappa
\newtheorem{theorem}{Theorem}[section]

\newtheorem{lemma}[theorem]{Lemma}

\def\AA{{\alpha }}

\def\AAJ{{\alpha ^\J}}
\def\AAmJ{{1-\alpha ^\J }}

\def\AAQ{{\alpha ^\Q}}

\def\AAJP{\alpha^\J}

\def\AAP{{\alpha}}
\def\AAQP{\alpha^\Q}
\def\eref#1{(\ref{#1})}
\def\HALF{{\textstyle\frac{1}{2}}}
\def\fract#1#2{{\textstyle\frac{#1}{#2}}}

\newcommand{\eL}[1]{\varepsilon_{\L,#1}}
\newcommand{\eR}[1]{\varepsilon_{\R,#1}}

\def\d{\partial}
\def\N{N}

\def\R{{\rm R}}
\def\L{{\rm L}}

\def\jL#1{j_{\L,#1}}
\def\jR#1{j_{\R,#1}}

\def\epsilon{\varepsilon}

\begin{document}

\title{Memory Effects in Nonequilibrium Transport for Deterministic Hamiltonian Systems}
\author{Jean-Pierre Eckmann${}^{1,2}$,
 Carlos~Mej\'\i{a}-Monasterio${}^1$ and Emmanuel Zabey${}^1$
}
\institute{${}^1$D\'epartement de Physique Th\'eorique, Universit\'e de
Gen\`eve\\${}^2$Section de Math\'ematiques, Universit\'e de
Gen\`eve\\}
\maketitle
\begin{abstract}
  We  consider  nonequilibrium  transport  in  a  simple  chain  of  identical
  mechanical cells  in which particles move  around. In each cell,  there is a
  rotating disc,  with which  these particles interact,  and this is  the only
  interaction in  the model.  It  was shown in \cite{eckmann-young}  that when
  the cells  are weakly coupled,  to a good  approximation, the jump  rates of
  particles  and the  energy-exchange rates  from cell  to cell  follow linear
  profiles.   Here, we  refine that  study by  analyzing  higher-order effects
  which are  induced by the presence  of external gradients  for situations in
  which memory effects, typical of  Hamiltonian dynamics, cannot be neglected. 
  For the steady state we propose  a set of balance equations for the particle
  number and energy  in terms of the reflection probabilities  of the cell and
  solve it  phenomenologically.  Using this approximate theory  we explain how
  these asymmetries affect  various aspects of heat and  particle transport in
  systems  of the  general type  described above  and obtain  in  the infinite
  volume  limit  the deviation  from  the  theory  in \cite{eckmann-young}  to
  first-order.    We   verify  our   assumptions   with  extensive   numerical
  simulations.
\end{abstract}
\thispagestyle{empty}

\section{Introduction}
\label{sec:intro}

In this paper, we study nonequilibrium transport in a class of 1-D Hamiltonian
systems consisting of  free noninteracting point particles of  mass $m$ moving
inside a chain of identical cells, which are like ``chaotic billiards''.  Each
cell  contains a device  (called a  \emph{tank} in  \cite{eckmann-young}) that
interacts with the particles by  exchanging energy with them.  Even though the
particles  do  not interact  among  themselves,  an  effective interaction  is
mediated   by    the   tanks.    As   pointed   out    in   previous   studies
\cite{Larralde2002,Larralde2003},   this   mediated   interaction  among   the
particles  allows  such  models  to  reach thermalization.  These  models  are
therefore    genuine    many-particle    interacting    Hamiltonian    systems
\cite{cohen-rondoni}.  Moreover, the details  of the interaction between tanks
and particles  are not important.   However, since our  aim is to  study model
systems with realistic microscopic dynamics we shall ask that this interaction
satisfies some  general conditions: The  system has to be  time-reversible and
Hamiltonian, although conservation of phase space volume is seemingly enough.

An  earlier study,  \cite{eckmann-young},  went some  ways  in explaining  the
energy  and particle  profiles in  terms of  a stochastic  approximation  to a
Hamiltonian  model. This  class  of models  was  derived, in  turn, from  work
\cite{Larralde2002,Larralde2003} in  which a  Lorentz gas with  rotating discs
was considered.

The  stochastic approximation  used  in \cite{eckmann-young}  assumed that  in
their  evolution, the  particles exit  on both  sides of  each cell  with {\em
  equal} probability at  some fixed rate.  Out of  equilibrium, the exit rates
will change  from cell to  cell and this  leads to the effective  transport of
particles and heat.   It was found that, to lowest order,  due to the gradient
character  \cite{spohn} of the  system, the  profiles for  the rates  at which
particles and energy are  transported among neighboring cells interpolate {\em
  linearly} between  the values imposed by  the baths at the  ends. From this,
energy and particle density along the chain were computed explicitly. They are
generally not linear.

The equal probability  to jump to the left  or to the right leads  to a simple
random walk process,  without any memory effects. This  assumption is strictly
valid only  when the cells are  weakly coupled, \emph{i.e.}, when  the size of
the openings connecting neighbouring cells  is very small.  However, away from
this limit  case, the memory  effects cannot in  general be neglected  and the
symmetry assumption is likely to fail.

In this paper,  we study the consequences of  including these dynamical memory
effects. The price  we pay for doing  this is the lack of  a simple stochastic
formulation to  describe the energy exchange  from cell to  cell.  Instead, at
the steady state we propose a set of balance equations for the particle number
and  energy  in  each  cell  in  terms  of  the  reflection  and  transmission
probabilities for particles and energy.  We specify a phenomenological law for
these probabilities that allows us to obtain an approximate expression for the
steady state.

Our findings for the class of $1$-D Hamiltonian chains that we consider can be
summarized as follows:
\begin{enumerate}
\item{}We  elaborate   a  phenomenological  theory  for   the  reflection  and
  transmission probabilities for particles and energy.
\item{}Close  to  equilibrium,  one   can  model  the  Hamiltonian  system  by
  persistent random  walks. In this case  the system remains  of gradient type
  and  thus,  the theory  of  \cite{eckmann-young}  is  still applicable  with
  corrections of the order of $1/N$ with $N$ being the size of the system.
\item{}Far  from   equilibrium,  we  demonstrate   that  there  is   indeed  a
  non-negligible  dependence on  the  local gradients,  and the  corresponding
  profiles can be  shown to obey a non-linear  differential equation that goes
  beyond linear response regime.
\end{enumerate}

For  our systems  this asymmetry  can be  seen as  a result  of  two different
contributions:  one  which is  purely  geometrical  and  another which  has  a
dynamical  origin.  For  any scattering  billiard (like  the model  cells) the
reflection  probability will  in general  be different  from  the transmission
purely due to the geometry of the cell.  When some source of interaction among
the particles is considered, these coefficients will depend in addition on the
local  thermodynamical  fields  (namely,   density  and  temperature)  of  the
particles  in the  cell.   When, in  addition,  the cell  is  subjected to  an
external   thermodynamical   gradient,   the   reflection   and   transmission
probabilities will be different  at the left and at the right  of the cell. In
this situation the dynamical asymmetry is emphasized.

Close to equilibrium, the geometric component dominates the dynamics, and then
the asymmetry  is \emph{independent} of  the local thermodynamical  fields and
thus, uniform along the chain.  Furthermore, in this simpler case, the bias of
the reflection/transmission  coefficients is the  {\em same} on both  sides of
any  cell. The  corresponding stochastic  model  is a  persistent random  walk
process  like  that  studied   in  Lorentz  gases  \cite{vanbeijeren}.   After
introducing     the     model     in     Section~\ref{sec:hamiltonian},     in
Section~\ref{sec:stochastic} we  sketch the stochastic  approximation on which
the   results   of   the    subsequent   sections   are   inspired   on.    In
Section~\ref{sec:profiles}, we  study the case of constant  asymmetry and show
that  one recovers  the  gradient property  observed in  \cite{eckmann-young},
albeit with a slight correction in the slope of the jump rate profiles.

In the final sections, we analyze  how the presence of {\it non uniform} local
fields changes this  picture.  Clearly, since the tanks  induce an interaction
of the particles, the reflection  (or transmission) probability depends on the
local thermodynamical  fields.  We will  distinguish two contributions  to the
dynamical asymmetry:  one that depends  on the mean  values of the  fields and
another one  that depends  on the  local gradients of  the fields.   These two
effects are added to the  purely geometrical contribution of the corresponding
non-interacting system.

In order  to judge the importance  of the asymmetry phenomenon,  we take fixed
external fields and let the number of cells be sufficiently large. This is the
standard limit  taken in studies  of the Fourier law  \cite{bonetto,lepri}. In
our  setting,  the  system is  no  longer  of  gradient  type, and  we  obtain
expressions for  the transport equations  which go beyond the  standard linear
regime.  While these equations remain approximate, we verify numerically their
validity even very far from equilibrium.

\section{The Mechanical Model}
\label{sec:hamiltonian}

We consider  a gas  of noninteracting  point particles of  mass $m$  that move
freely   inside   a   one-dimensional   chain  composed   of   $N$   identical
two-dimensional cells,  arranged horizontally. Each  cell is connected  to its
left and right neighbors through two openings of size $\gamma$.

Inside each cell  there is a mechanical ``tank'' capable  of storing a certain
amount of  energy. When a particle  hits the tank, some  energy exchange takes
place, and this is the only interaction in the model.

\begin{figure}[ht]
\begin{center}
\includegraphics[scale=0.65]{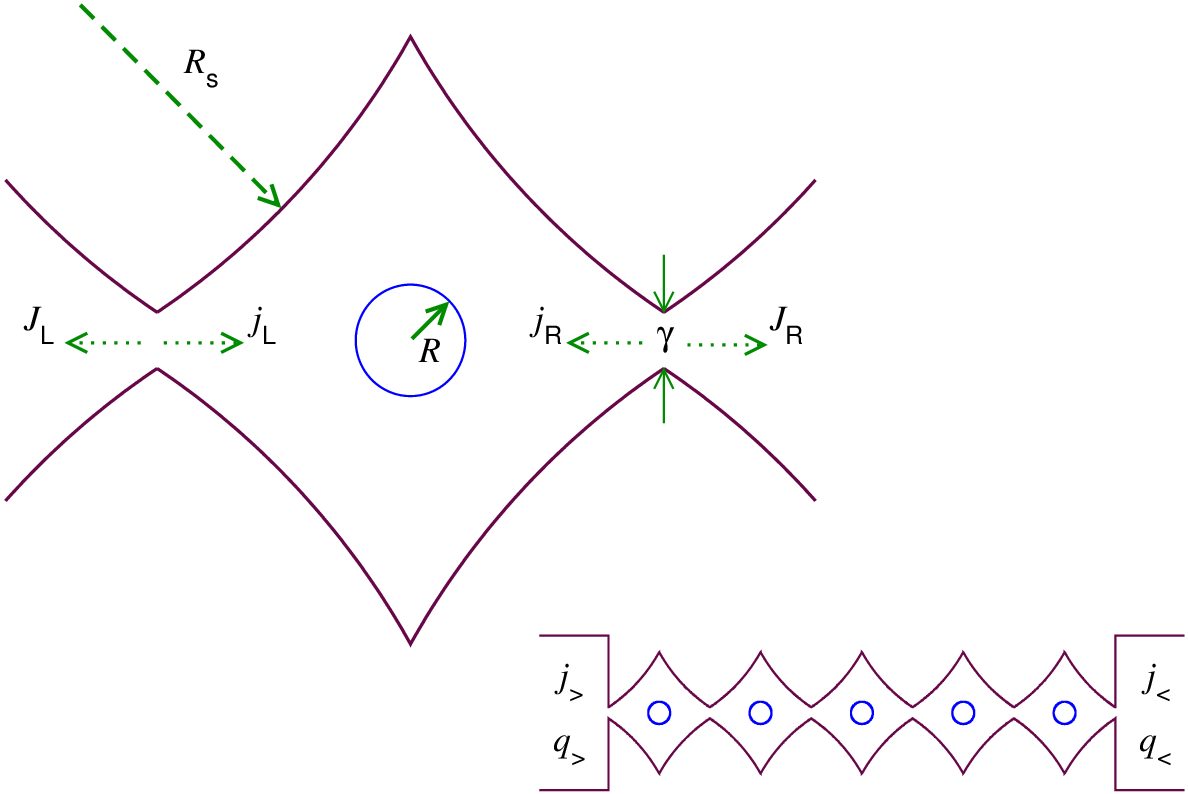}
\caption{
  Upper left: The  geometry of a single cell for the  rotating disc model. The
  boundary of the  cell is made of  four arcs of circles of  radius $R_s$. The
  radius $R$  of the rotating disc is  chosen so that no  trajectory can cross
  the cell without undergoing any collision. The arrows represent the incoming
  ($j$) and  outgoing ($J$) rates of  particles that cross the  exits of width
  $\gamma$. The rates $q$ and $Q$ for the energy are not shown. Lower right: A
  chain of 5 cells connected to two baths.
 \label{fig:geometry}}
\end{center}
\end{figure}

While  the  details  of  this  interaction  are  largely  irrelevant  for  the
discussion in  this paper,  our simulations have  been done for  the following
precise setup which we call the  rotating disc model (RDM).  Each cell has the
geometry described in Fig.~\ref{fig:geometry}. The energy tank is modeled by a
freely rotating disc which is pinned at  the center of the cell. Its energy is
only rotational.  When  a particle hits the disc,  it exchanges its tangential
velocity $v_t$  with the  angular velocity  $\omega $ of  the disc,  while the
normal component $v_n$ is reflected \cite{Larralde2002}:
\begin{equ}\label{eq:coll-rule}
 \omega '=v_t~,\quad
 v_t'=\omega ~,\quad
 v_n'=-v_n~.
\end{equ}
Note that while the particles do not see each other, they effectively interact
through their collisions with the tanks. Thus local equilibrium can be reached
\cite{dhar}. The model is by no means free.

The chain  is connected at  both ends to  two reservoirs of  particles through
openings of the  same size $\gamma$. The reservoirs  are idealized as infinite
chambers containing an ideal gas at a certain density $n$ and temperature $T$.
When we study non-equilibrium effects, $n$ and $T$ will differ at both ends.

Another  variant of  this model  has the  same fixed  walls, but  the  disc is
replaced by a rotating needle, with elastic collision, conserving total energy
(and total angular momentum).

These  models are among  a class  of models  satisfying the  following minimal
assumptions:
\begin{enumerate}
\item[$\bullet$] \emph{Time  reversibility}: The  trajectory of the  system in
  phase space across  a scattering event must be  time reversible.  While this
  condition is not  really necessary for our derivations, our  aim is to model
  realistic macroscopic situations for which microscopic time-reversibility is
  believed to hold.
\item[$\bullet$] \emph{Conservation of  phase-space volume}: This condition is
  not as strong as requiring the dynamics to be Hamiltonian, and seems to lead
  to similar  results. In particular,  the rotating disc model  satisfies this
  condition but is  not Hamiltonian: Its collision rule  preserves phase space
  volume but is not a canonical transformation.
\end{enumerate}
We call the models satisfying these two assumptions {\bf mechanical}.

We end the description of the model by a more detailed account of the heat and
particle  baths.  From  the left  reservoir at  density $n_>$  and temperature
$T_>$, particles are  injected into the system at a  rate $\j_>$. Observe that
the mean  energy of the  particles injected into  the system is not  $T_>$ but
$\fract{3}{2}T_>$  (see  {\it e.g.},  \cite{eckmann-young}).   Thus, the  left
reservoir injects energy into the system at a rate $q_>$ given by
\begin{equ}
q_>=\frac32\,j_>\,T_>~.
\end{equ}
Analogously, the right  reservoir injects particles and energy  at rates $j_<$
and $q_<$ respectively.  Each injected particle will be eventually re-absorbed
by one of the reservoirs and this happens when it crosses into that reservoir.
All information about that particle is then discarded.

As  the reservoirs  consist  of an  ideal gas,  the  injection rate  $j$ is  a
function of the density $n$ and temperature of the reservoir given by
\begin{equ} \label{eq:injection}
j \propto \gamma n T^{1/2} \ .
\end{equ}
Furthermore, the chemical potential at  the reservoirs can be written in terms
of the injection rate as
\begin{equ} \label{eq:chem-pot}
\mu = T\log \left(\frac{\lambda_0 j}{T^{3/2}}\right)
\end{equ}
with $\lambda_0$ some constant. 

We end the section  with some comments on the choice of  the model: Our choice
of the  reflecting boundaries of  Fig.~\ref{fig:geometry} is to  guarantee the
desirable ergodic and mixing properties of chaotic billiards. However, we have
not succeeded in showing rigorously that the cell is either ergodic or mixing.
While the geometry  of the rotating needles model (RNM) is  similar to that of
the RDM, there are two main differences: First, the RNM is not only mechanical
but Hamiltonian. Second,  in the RNM model, a particle  can in principle cross
many  cells  without hitting  any  needles  or  boundaries. This  may  produce
logarithmic corrections to transport, which are beyond our study.

\section{An Approximate Stochastic Model}
\label{sec:stochastic}

In \cite{eckmann-young},  a stochastic  approximation of the  mechanical model
was considered. In this approximation it  was assumed that what happens in one
cell is independent of the state  of the other cells.  The particles perform a
random walk among the cells, mixing their energies with the tanks according to
rules which mimic the mechanical system described above. Once the particle has
mixed its energy,  it waits (until an exponential clock  rings) and then jumps
with equal probability to the left or to the right.  With these assumptions it
was found that the rates at  which particles and energy jump among neighboring
cells are linear functions of the  position and from this, energy and particle
density  along  the  chain  can  be computed  explicitly  under  very  general
assumptions.  However,  note that these  assumptions are valid when  the cells
are weakly  coupled and dynamical memory  effects appear when this  is not the
case.

The purpose of this paper is to study the consequences of these memory effects
on the  transport properties of the  mechanical models that we  consider here. 
The price we pay is that the stochastic model, used in \cite{eckmann-young} is
not longer  appropriate. We  briefly sketch the  essentials of  this stochastic
approximation that will serve us as  a guideline to our analysis (we refer the
interested reader to \cite{eckmann-young} for more details about this).

The stochastic description  is based on particles, carrying  energies, and the
discs (as in \cite{eckmann-young}).  Each  state of the cell is represented by
a         point        in        $\Omega=\cup_m         \Omega^m$        where
$\Omega_m=\{(x_1,\dots,x_m),(p_1,\dots,p_m),y\}$  and  $m$  is the  number  of
particles in the  cell at some given time, $x_i$ are  their energies, $p_i \in
\{-1,1\}$ specifies the  side on which the particle enter the  cell and $y$ is
the energy stored by the disc.  For a chain with $N$ cells, the phase space is
$\Omega^{(N)}=\Omega\times\Omega\times\cdots\times\Omega$  with  $N$ factors.  
The reservoirs  at the ends absorb particles,  and eject them at  a rate $j_>$
resp.   $j_<$ (for left  and right)  and with  an exponential  distribution of
energy (with mean temperatures $T_>$ and $T_<$).

The  stochastic  process is  as  follows  (see  \cite{eckmann-young} for  more
details): When a particle  entered a cell $k$ from the left,  it will wait for
an exponentially  distributed time (whose mean  may depend on  its energy), it
mixes energy  with the  disc, and leaves  the cell (after  another exponential
waiting  time)  with probability  $\alpha_{\L,k}^\J$  to  the  left, and  with
probability $1-\alpha _{\L,k}^\J$  to the right.

This  stochastic  description leads,  in  the  steady  state, to  the  balance
equations for rates of particle injection (see \eref{eq:balance} below).

However,  while this  description is  adequate for  particle flux,  it  is too
detailed for  the energy  flux.  Instead of  trying to formulate  a stochastic
process  for both  particle  and  energy fluxes  (see  \cite{uchiyama} for  an
example), we  elaborate a phenomenological theory for  the physical parameters
involved.  We show  that this information is enough  to obtain precise results
on the average properties of the steady state of this type of systems.

It  is convenient  to distinguish  the  local \emph{incoming}  rates from  the
\emph{outgoing} rates.  We denote by $J_{\L,k}  $ and $Q_{\L,k} $ the rates at
which particles and  energy exit the $k$-th cell to the  left and by $J_{\R,k}
$, $Q_{\R,k} $  those to the right. The exit rates  are then simply determined
by the  corresponding incoming rates  $j_{\L/\R,k}$ and $q_{\L/\R,k}$,  and by
reflection  $\alpha_{\L,k}$, $\alpha_{\R,k}$  coefficients which  will satisfy
the balance equations (for the cell $k$):
\begin{equa}[2] \label{eq:balance}
&J_{\L,k} & \ = \ & \alpha^\J_{\L,k}~j_{\L,k} + (1-\alpha^\J_{\R,k})~j_{\R,k} \ ,\\
&J_{\R,k} & \ = \ & (1-\alpha^\J_{\L,k})~j_{\L,k} + \alpha^\J_{\R,k}~j_{\R,k} \ ,\\
\\
&Q_{\L,k} & \ = \ & \alpha^\Q_{\L,k}~q_{\L,k} + (1-\alpha^\Q_{\R,k})~q_{\R,k} \ ,\\
&Q_{\R,k} & \ = \ & (1-\alpha^\Q_{\L,k})~q_{\L,k} + \alpha^\Q_{\R,k}~q_{\R,k} \ .\\
\end{equa}
Note  that even  for  (left-right)  symmetric cells,  out  of equilibrium  the
reflection coefficients at the left and  right of the cell are not necessarily
equal.   The balance  equations account  for  the conservation  of energy  and
particle    number,     namely    $Q_{\L,k}+Q_{\R,k}=q_{\L,k}+q_{\R,k}$    and
$J_{\L,k}+J_{\R,k}=j_{\L,k}+j_{\R,k}$.  Since  an average the  steady state is
assumed, mean rates appear in \eref{eq:balance}.

We  define the  coefficients by  empirical probabilities  $j_{\L\R}~\ldots$ by
(omitting the index $k$ for better legibility)
\begin{equa}[5]\label{eq:better}
  \alpha^\J_\L&=&\frac{j_{\L\L}}{j_\L}~,\quad 
  \alpha^\J_\R&=&\frac{j_{\R\R}}{j_\R}~,\\
  \alpha^\Q_\L&=&\frac{1}{2}+\frac{q_{\L\L}-q_{\L\R}}{2q_\L}~,\quad
  \alpha^\Q_\R&=&\frac{1}{2}+\frac{q_{\R\R}-q_{\R\L}}{2q_\R}~,
\end{equa}
where, for example, $j_{\L\R}$ is the  rate of particles leaving on the right,
which entered  on the left, and $q_{\L\R}$  is the mean energy  carried out by
particles which  entered on the left.  If we denote  $p(E'|E)$ the conditional
probability that a  particle entering with energy $E$  leaves with energy $E'$
and by $\alpha _{\L\L}(E',E)$ the probability  that it leaves on the left when
it entered on the left, then one can think of $q_{\L\L}$ as
\begin{equ}
q_{\L\L} = \int dE' \,dE \,E' \alpha _{\L\L}(E',E)\, p(E'|E) ~,
\end{equ}
where the complicated dependence on the state of the system is neglected (that
is, we do not  consider the density and the other particles  in this formula). 
Instead  we  will  use  a  phenomenological  description  for  the  reflection
coefficients.  Perhaps it is useful to note that with the above conventions,
\begin{equ}
j_{\L\L} = \int dE' \,dE \, \alpha _{\L\L}(E',E)\, p(E'|E) ~.
\end{equ}

The definition for the $\alpha ^\J$  is canonical, but for the $\alpha ^\Q$ we
chose a more complicated  expression: It simultaneously preserves total energy
conservation, but allows  for an energy change during  scattering.  The reader
should note that  the quantities $\alpha $, $J$, $Q$  are mean values averaged
on the (fluctuating) steady state. An  important aspect of the present work is
to check that this approximation still captures the essentials of transport of
heat and particles.

To further  simplify the discussion,  we make the approximate  assumption that
the $\alpha^X_Y$ (for all $X\in \{\J,\Q\}$ and $Y\in\{\L,\R\}$) only depend on
the incoming  fluxes.  Due to the  mechanical nature of  the models considered
here, the time scale is a free variable leading to the scaling relation
\begin{equation}\label{eq:scaling}
\alpha (j_\L,j_\R,q_\L,q_\R) \,=\,\alpha (\lambda j_\L,\lambda
j_\R,\lambda ^3q_\L,\lambda ^3q_\R)~,
\end{equation}
for all $\lambda >0$. Therefore, $\alpha$ only depends on 3 ratios. It will be
useful  to  distinguish  the  contribution  to $\alpha$  that  arises  out  of
equilibrium from the contribution that only  depends on the mean values of the
fields.  Accordingly we write
\begin{equ} \label{eq:alpha-cont}
 \alpha (j_\L,j_\R,q_\L,q_\R) = \alpha _\G\left(\frac{j^{3/2}}{q^{1/2}}\right)
 +  \epsilon  \left (\frac{j^{3/2}}{q^{1/2}},\frac{j_\R-j_\L}{j_\R+j_\L},\frac
 {q_\R-q_\L}{q_\R+q_\L}\right ) \ ,
\end{equ}
and  require   that  $\epsilon  (j^{3/2}/q^{1/2},0,0)=0$,   \emph{i.e.},  that
$\epsilon$ vanishes at equilibrium. The  term $\alpha _\G$, which has a purely
geometric  origin, describes  in turns  those  aspects which  hold at  thermal
equilibrium  and  the  term  $j^{3/2}/q^{1/2}$, with  $  q=(q_\L+q_\R)/2$  and
$j=(j_\L+j_\R)/2$,     is     proportional     to     the     density     (see
Eq.~\eref{eq:injection}).

Note that  $\alpha^X_Y=\HALF$ corresponds to the simple  symmetric random walk
considered  in  \cite{eckmann-young}.   The  case when  the  $\alpha^X_Y$  are
independent of the  local fields, but different from  $\HALF$ corresponds to a
persistent random walk and will be  studied in the next section.  More general
and realistic laws will be investigated in the last sections.

Also note that the details of the dynamics, in particular those that depend on
the  specific model,  are  encoded by  Eq.~\ref{eq:alpha-cont}.   To show  the
validity of  our approach we will  proceed as follows: We  first introduce the
law \eref{eq:alpha-cont}.   This closes the  balance equations for  the energy
and  particle number.   We  next determine  the  thermodynamical profiles  and
currents.  Finally we compare these ``theoretical'' profiles to those obtained
with numerical simulations of the mechanical model.

\section{Solution of the Stochastic Model Near Equilibrium}
\label{sec:profiles}

In this section,  we discuss profiles in the infinite  volume limit with fixed
boundary conditions, under the approximating  assumption that every cell is at
local  equilibrium.   This  means  that  we neglect  the  term  $\epsilon$  in
Eq.~\eref{eq:alpha-cont} and  thus, $\alpha_{\R,k}^\J=\alpha_{\L,k}^\J =\alpha
_\G^\J$  and  similarly  for  the  $\alpha ^\Q$.   Furthermore,  we  make  the
assumption, that  $\alpha _\G^\J$ and  $\alpha _\G^\Q$ are independent  of the
densities found in the chain. In this approximation the reflection coefficient
is a constant independent of  the thermodynamical fields and thus, independent
of position.  The exact range of validity of this assumption will be discussed
in Sec.~\ref{sec:bias}. Our  approximations mean that we are  dealing with the
case  of   a  persistent   random  walk,  and   the  current   section,  while
``well-known'' in  principle, serves as a  first check of the  validity of our
approximations. Once  this check has been  done, we will study  in more detail
the corrections to the persistent case.

Since   the   geometric   contribution   is   the  only   one   remaining   in
(\ref{eq:alpha-cont}),  we   refer  to  this   case  as  the   {\bf  geometric
  approximation}.   Note that  this contribution  will  lead in  general to  a
reflection coefficient different  from $\HALF$. This approximation corresponds
to a  persistent random walk in which  the probability for the  walker to move
forward is different from the probability to move backwards.

We   discuss   the    total   ejection   rates   $J_k=J_{\L,k}+J_{\R,k}$   and
$Q_k=Q_{\L,k}+Q_{\R,k}$. The boundary conditions of the problem are simply
\begin{equ}[eq:ident]
j_>=j_{\L,1}~,\quad
j_<=j_{\R,N},\quad 
q_>=q_{\L,1}~,\quad
q_<=q_{\R,N}~.
\end{equ}
From  now on  we will  use rescaled  variables $\xi=k/(N+1)\in[0,1]$,  and let
$J(\xi)=J_k$ and $Q(\xi)=Q_k$. For the  sake of simplicity, we will only write
$\alpha^\J$ and $\alpha^\Q$.

\begin{lemma}\label{lemma:1}
  Under the  assumptions made  above, the particle  and energy  ejection rates
  satisfy
\begin{equ} \label{eq:prof_j}
J(\xi) = 2(j_> + \xi\cdot\Delta j) +
\frac{(1-2\AAJP)(1-2\xi)}{1+(N-1)\AAJP}\Delta j~,
\end{equ}
where $\Delta j = j_< - j_> $, and
\begin{equ} \label{eq:prof_q}
Q(\xi) = 2(q_> + \xi\cdot\Delta q) +
\frac{(1-2\AAQP)(1-2\xi)}{1+(N-1)\AAQP}
\Delta q\ ,
\end{equ}
with $\Delta q = q_<-q_> $.
\end{lemma}

\noindent{\bf Remarks}:
The   first   term   in   the   \emph{r.h.s.}~of   Eqs.~\eref{eq:prof_j}   and
\eref{eq:prof_q} corresponds  to the result obtained  in \cite{eckmann-young}. 
The  last terms  account  for the  correction  when $\AA  \ne  \HALF $.  These
corrections disappear in the infinite volume limit. The integral over $\xi$ of
the correction  vanishes, because of energy and  particle conservation.  Also,
note that as  $\AA$ is a constant, the profiles are  linear (even when $\alpha
\ne\HALF$), because the system is still of gradient type.

\noindent{\bf Proof}:
The proof is  just a calculation. We will concentrate on  the case of particle
rates.  This calculation starts  from \eref{eq:balance}.  Using the identities
Eq.\eref{eq:ident}, we  can solve \eref{eq:balance} for the  remaining $J$ and
obtain for $k=1,\dots,N$,
\begin{equa}[4] \label{eq:sol_j}
J_{\L,k} & = & \frac{(N-k+1)\alpha}{1+(N-1)\alpha}j_> &+&
\frac{1+(k-2)\alpha}{1+(N-1)\alpha}j_< ~, \\
J_{\R,k} & = & \frac{1+(N-k-1)\alpha}{1+(N-1)\alpha}j_> &+& 
\frac{k\alpha}{1+(N-1)\alpha}j_< \ .\\
\end{equa}
Then, summing the $\R$ and $\L$ terms, we get
\begin{equ} \label{eq:jump_j}
J_k = \frac{1+2(N-k)\AA}{1+(N-1)\AA}j_> +
\frac{1+2(k-1)\AA}{1+(N-1)\AA}j_< ~.
\end{equ}
Rearranging  terms   one  immediately  obtains   \eref{eq:prof_j}.   The  case
\eref{eq:prof_q} is handled similarly.

Therefore, if the value for $\AAP$ is known, the profile of the ejection rates
can  be obtained  from  the solution  \eref{eq:sol_j}  for any  nonequilibrium
state, as long as $\AAP$ is independent of the external parameters.

\def\nzero{\eta_0}

\subsection{Thermodynamical profiles}

In order to  identify temperature and particle density,  we use the techniques
of \cite{eckmann-young}. In that paper, it was observed that in equilibrium, a
single cell has a Poisson distribution  of particle number with mean $n$ and a
canonical distribution  of the total energy  in the cell  (particles and disc)
with   a   temperature   $T$.     Furthermore,   (see   Proposition   4.1   in
\cite{eckmann-young}) one has universal relations between these quantities and
the properties of the reservoir, namely
\begin{equ} 
T \ = \ \frac{2}{3} \ \frac{Q}{J} ~;
\end{equ}
\begin{equ} 
n \ = \ \nzero\frac{\rate^{3/2}}{ {Q^{1/2}}}~.
\end{equ}
The constant $\nzero$ depends only on the geometry of the cell and the size of
the opening, but not on the shape
\begin{equ}\label{eq:n-0}
\nzero= \sqrt{\frac{3\pi}{4}}\,\frac{\text{Area}(\Gamma)}{|\gamma|}~,
\end{equ}
where the area is that of the  cell minus the disc and where $|\gamma|$ is the
size  of the  opening between  adjacent cells.  This identification  is unique
provided the  system is  ergodic.\footnote{The definition of  $\nzero$ differs
  from that of \cite{eckmann-young} because of different normalizations of the
  kinetic energy.}

Assuming local equilibrium  at any cell $k$ in the  system, the formulas above
generalize  immediately to  predictions of  the profiles  for  temperature and
particle number (with $\xi=k/(N+1)$)
\begin{equ} \label{eq:T}
T(\xi) \ = \ \frac{2}{3} \ \frac{Q(\xi)}{J(\xi)} ~;
\end{equ}
\begin{equ} \label{eq:kappa}
n(\xi) \ = \ \nzero\frac{\rate^{3/2}(\xi)}{ {Q^{1/2}(\xi)}}~.
\end{equ}
Note that while the jump rates $Q$ and $J$ have linear profiles in the current
approximation  the  profiles  of  $T$  and $n$  are  generally  nonlinear.  In
Sect.~\ref{sec:asymmetry}  we  will  study   in  detail  the  deviations  from
linearity of $Q$ and $J$.

To  test  these  results  we  performed  out of  (but  close  to)  equilibrium
simulations for a chain of 20 RDM cells with $|\gamma|=0.08$, $R_s=1.15$ and a
disc radius of $R=0.0793$ (ensuring that  no particle can cross a cell without
being  scattered).   The choice  parameters  used  in  the simulations  are  a
compromise between good mixing within  each cell, and the speed of convergence
to the stationary state.

\begin{figure}[!t]
\begin{center}
\includegraphics[scale=0.55]{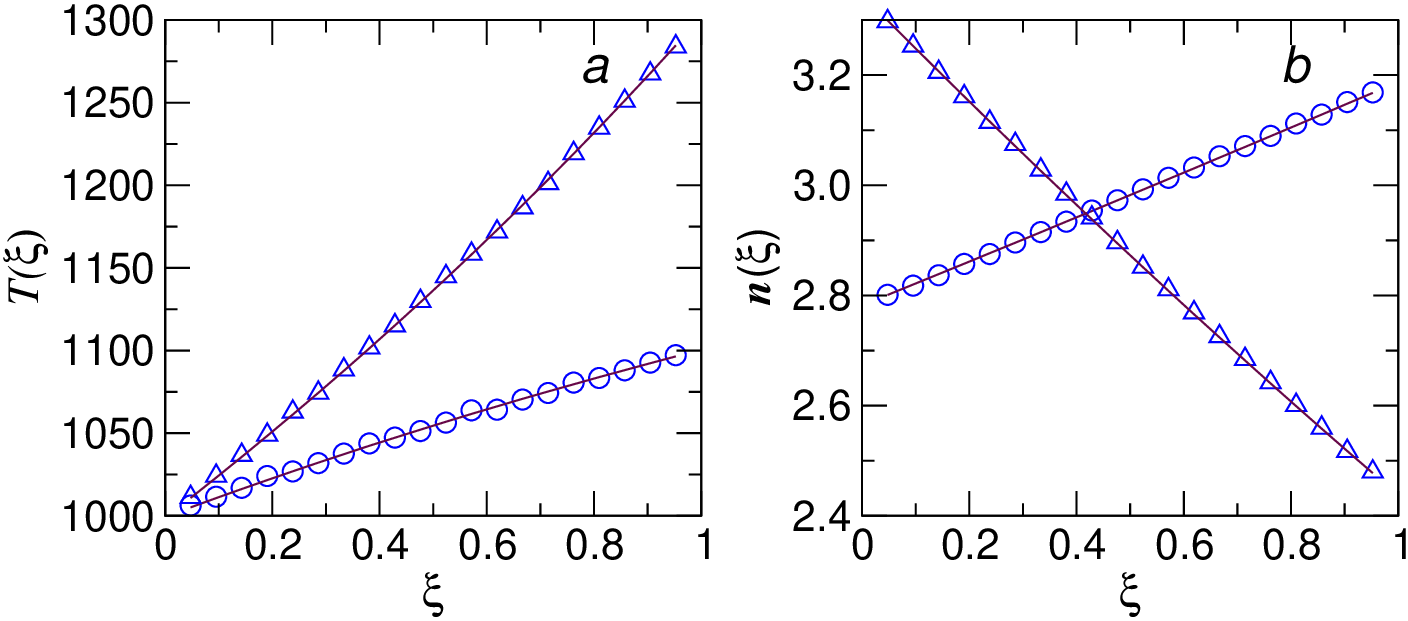}
\caption{
  Comparison   of  the  measured   profiles  with   those  predicted   by  the
  Eqs.~\eref{eq:T}            and            \eref{eq:kappa},            using
  \eref{eq:prof_j}-\eref{eq:prof_q}.  The  measured profiles were  obtained as
  an average over  400 different realizations of the RDM  for $20$ cells.  The
  energy  profile $T(\xi)$  ($a$) and  the particle  density  profile $n(\xi)$
  ($b$)  are shown  for  two different  simulations  with $j_>=10$,  $j_<=12$,
  $T_>=1000$ and  $T_<=1100$ (circles) and $j_>=12$,  $j_<=10$, $T_>=1000$ and
  $T_<=1300$ (triangles).   For all data the  error bars are  smaller than the
  symbol size.   The solid  lines in each  panel correspond to  the analytical
  profiles  of  Eqs.~\eref{eq:T}  and  \eref{eq:kappa} for  $\AAJ=0.5585$  and
  $\AAQ=0.5609$ as explained in the text.
\label{fig:th-profiles}}
\end{center}
\end{figure}

We checked in  our simulations for the Hamiltonian model  that the profiles of
the outgoing rates are described by \eref{eq:prof_j} and \eref{eq:prof_q}.  In
Fig.~\ref{fig:th-profiles} we  show the  temperature and density  profiles for
two different simulations, in which  the baths were set to $j_>=10$, $j_<=12$,
$T_>=1000$ and  $T_<=1100$ (circles),  and $j_>=12$, $j_<=10$,  $T_>=1000$ and
$T_<=1300$ (triangles).

To  compare  the  measured  profiles  for  $T$ and  $n$  with  our  analytical
expressions  we have  first numerically  computed the  reflection coefficients
$\AAJ$ and  $\AAQ$ for  each cell.   For both simulations  we found  that both
coefficients are constant (to within  numerical accuracy) along the chain with
a  mean value  of $\AAJ=0.5585\pm  10^{-4}$ and  $\AAQ=0.5609\pm  10^{-4}$. We
insert  these   values  into  \eref{eq:prof_j}--\eref{eq:prof_q}   and  obtain
estimates for  the profiles of the  outgoing rates $J(\xi)$  and $Q(\xi)$.  We
then  use these  estimates in  \eref{eq:T} and  \eref{eq:kappa} to  obtain the
profiles for the  energy $T$ and the number of particles  $n$, as predicted by
our theory.   These are  shown as solid  lines in  Fig.~\ref{fig:th-profiles}. 
The fit  with the  directly measured profiles  for $T$  and $n$ is  excellent. 
Note that these  profiles are not expected to  be linear \cite{eckmann-young}. 
This  non-linearity   has  already  been   verified  against  the   theory  in
\cite{eckmann-young}, but  here we study rather  the effect of  the asymmetry. 
At this order of approximation, that is, in the case of the persistent $\alpha
$'s the  difference between the two approaches  is only of order  $1/N$ and is
not yet visible in the simulations.

\subsection{Macroscopic currents}
 
The heat and matter local macroscopic  currents can be easily derived from the
balance equations.  The  current of particles between the  $k$-th and $k+1$-st
cells  is defined  as  $\varphi^\J =  J_{\R,k}  - j_{\R,k}$  or,  in terms  of
outgoing  jump   rates  as  $\varphi^\J  =  J_{\R,k}   -  J_{\L,k+1}$.   Using
Eq.~\eref{eq:sol_j} we obtain
\begin{equ} \label{eq:curr-j}
\varphi^\J  = -\frac{\AAmJ}{1+(N-1)\AAJ}\Delta j ~,
\end{equ}
where  the result  is of  course  independent of  $k$.  In  an analogous  way,
$\varphi^\Q = Q_{\R,k} - Q_{\L,k+1}$, and we obtain for the heat current
\begin{equ} \label{eq:curr-q}
\varphi^\Q   = -\frac{1-\AAQ}{1+(N-1)\AAQ}\Delta q \ .
\end{equ}
In  these  variables,  the  currents  are linear  functions  of  the  external
gradients.

In the  numerical simulations described above,  we also computed  the heat and
matter currents obtaining an average value of: $\varphi^\J = -0.077 \pm 0.002$
and  $\varphi^\Q   =  -185  \pm  13$   (for  the  experiment   in  circles  in
Fig.~\ref{fig:th-profiles}) and $\varphi^\J =  0.08 \pm 0.008$ and $\varphi^\Q
= -56 \pm 4$ (for the experiment in triangles of the same figure).  If instead
we  insert  the  numerically  computed  values  for  $\AAJ$  and  $\AAQ$  into
Eqs.~\eref{eq:curr-j}  and \eref{eq:curr-q} we  obtain $\varphi^\J  = -0.076$,
$\varphi^\Q = -180.8$ (circles) and $\varphi^\J = 0.076$, $\varphi^\Q = -56.5$
in  perfect  agreement  with  the  numerical experiment  to  within  numerical
accuracy.

Therefore, we  conclude that in  cases in which  the $\AA$ probability  can be
taken to be  constant along the chain, the  conclusions of Lemma~\ref{lemma:1}
describe remarkably well  the transport properties of the  disc model. Similar
results were found  for the needle model, and we  conjecture that this extends
to many similar models.

The  validity   of  the   linear  transport  equations   \eref{eq:curr-j}  and
\eref{eq:curr-q}  depends crucially  on the  property that  $\alpha _\L=\alpha
_\R$.  For  high   gradients,  this  property  is  violated   as  we  show  in
Sec.~\ref{sec:asymmetry}.  In  that case, the  Fourier law does not  hold, but
the flux  still scales as  $1/N$ for large  systems. The reason is  that local
gradients appear in the diffusion constants.

We remark finally that $\AA$ (and thus the details of the cell) determines the
nature of  the macroscopic  transport: If  $\AA = 0$  the transport  along the
chain is ballistic  as the currents do  not scale with the size  of the system
$N$. On the  other hand, if $\AA =  1$ the chain behaves as  an insulator. For
any other value  of $\AA$ the transport is normal  with well defined transport
coefficients.

\section{Validity of the Geometric Approximation}
\label{sec:bias}

In the  previous section,  we assumed that  the reflection  coefficient $\AA$,
while  not necessarily  equal  to  $\HALF$, neither  depends  on the  particle
density  inside a  cell  nor on  the strain  acting  on it.   In other  words,
$\alpha^\J_{\R,k}=\alpha   ^\J_{\L,k}=\alpha   ^\J_\G$,   this   value   being
determined by the geometry of the cell only (and similarly for $\AAQP$).

In  this  section  we discuss  this  assumption  and  determine the  range  of
parameters for which it applies  with high precision. The discussion will also
clarify the  choice of the decomposition  (\ref{eq:alpha-cont}).  The profiles
obtained when $\AA_\R$ and $\AA_\L$ are different and vary along the chain are
discussed in the next section.

The reader  should first  note a simple  fact: When  the discs (tanks)  in the
cells are  not allowed  to rotate,  and the reflection  is specular,  then the
trajectories are  independent of  the energy and  of the particle  density. In
that  case  $\AA$ is  clearly  independent  of  all external  parameters  (and
$\AAJP=\AAQP$ since  the particles are  the only energy  carriers). Therefore,
any variation  of $\AA$  has its  origin in the  effective interaction  of the
particles.

The  interaction changes the  dynamics inside  the cell,  but does  not always
induce a variation  of $\AA$.  We first consider an  ideal case: particles are
always trapped in  the cell for a  very long time. They collide  with the tank
many times, and forget  the side from which they entered the  cell, as well as
the energy  they had at  that time.  In  this case, the thermalization  of the
particles is perfect {\em due to the interaction} and no asymmetry appears.

The  geometric approximation  loses its  validity  due to  two related  memory
effects:
\begin{enumerate}
\item[$\bullet$] The  distribution of energy in  the cell is  not uniform, and
  thus thermalization is only approximate
\item[$\bullet$]  Particles  retain a  memory  of  the  energy they  had  when
  entering the cell, and this effect is energy-dependent.
\end{enumerate}

\begin{figure}[!t]
\begin{center}
\includegraphics[scale=0.6]{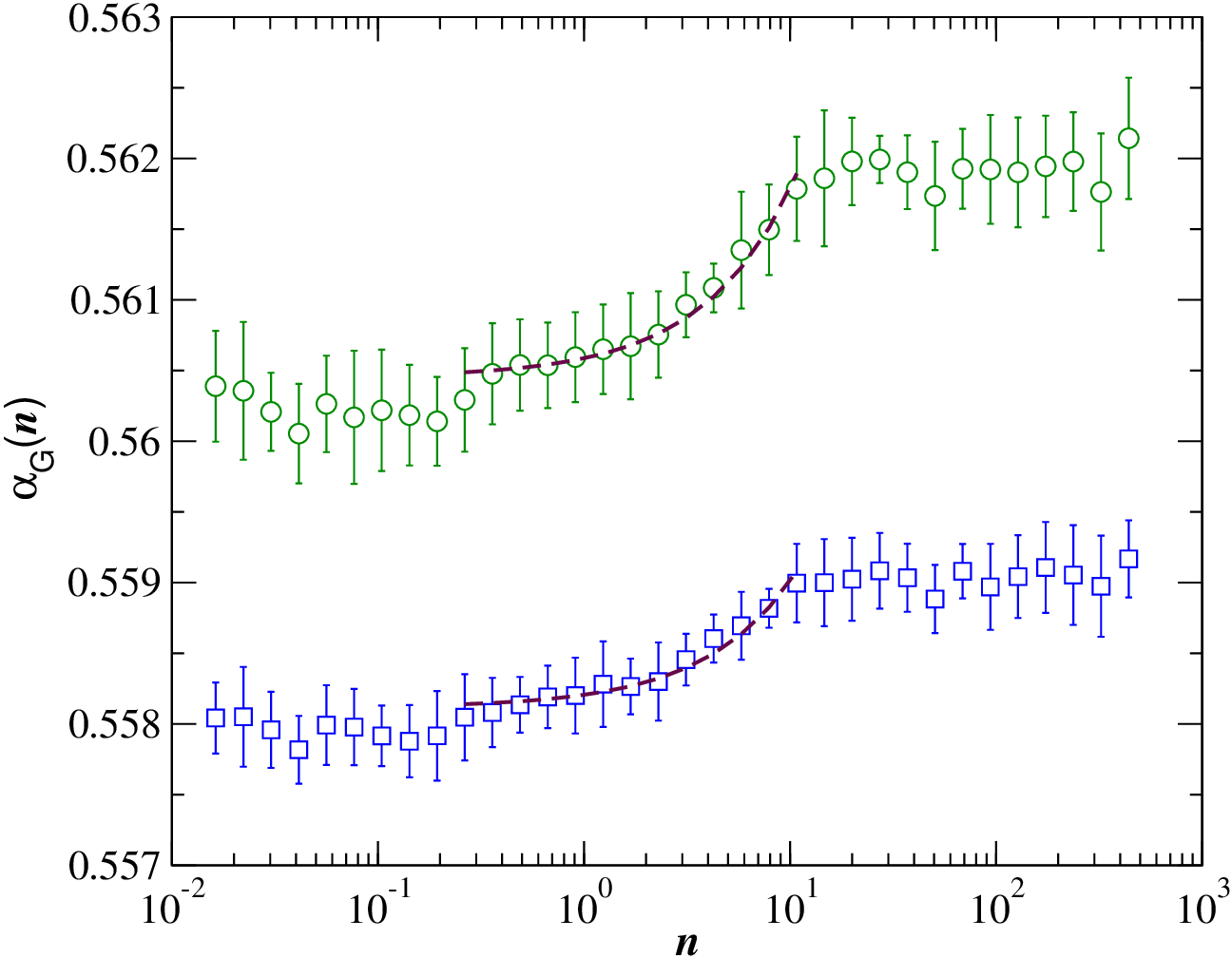}
\caption{Dependence of the equilibrium reflection probabilities $\AA^\J_\G$
  (squares) and $\AA^\Q_\G$ (circles) on  the density of particles in the cell
  $n$ for  the RDM  with $\gamma=0.08$. Note  that $\alpha_\G$ saturates  to a
  constant  value  at  low  and  high densities.   At  intermediate  densities
  $0.3<n<10$, $\alpha_\G$  grows linearly with density as  corroborated by the
  linear fits (dashed lines).  The fits are $\alpha _\G^\Q(n)=0.56048\pm2\cdot
  10^{-5}+(1.29\cdot    10^{-4}\pm     5\cdot    10^{-6})n$,    and    $\alpha
  _\G^\J(n)=0.55813\pm2\cdot 10^{-5}+(8.71\cdot 10^{-5}\pm 5\cdot 10^{-6})n$.
 \label{fig:n-dep}}
\end{center}
\end{figure}

In  order to  test  these  properties, we  first  performed {\em  equilibrium}
simulations.  These will isolate  the geometric  contributions to  the $\alpha
$'s.  This allows us,  in the  next section,  to identify  those contributions
which are typical of  non-equilibrium, justifying thereby the decomposition of
Eq.~\eref{eq:alpha-cont}.  In  Fig.~\ref{fig:n-dep} we show  the dependence of
the  reflection probabilities $\AA^\J_\G$  and $\AA^\Q_\G$  on the  density of
particles  $n$ for  a  single cell  (for  the RDM).   Injection  rates of  the
reservoirs are chosen according to  \eref{eq:injection} in order to obtain the
desired densities.  We observe the following:
\begin{itemize}  
\item For  low densities  ($n\lesssim0.3$) and high  densities ($n\gtrsim10$),
  the $\AA$ is practically constant.  Therefore,  if a whole chain of cells is
  in either of  these regimes, we can  model it with a constant  $\AA$.  If in
  addition,  the local  gradients are  small, then  the system  is  (close to)
  gradient   type,  and   is   well  described   by   the  approximations   of
  Sec.~\ref{sec:profiles}.
  
\item For intermediate  densities, $\AA$ increases linearly with  the density. 
  This is  a new kind of  regime (not of  gradient type), which we  discuss in
  more detail in the next section.
\end{itemize}

As    discussed    in    Sec.~\ref{sec:stochastic},   the    distinction    in
Eq.~\eref{eq:alpha-cont}  between the  geometric contribution  $\alpha_\G$ and
the dynamical $\varepsilon$, is that  while $\varepsilon$ depends on the local
gradients  of  the  thermodynamical   field  (and  is  zero  at  equilibrium),
$\alpha_\G$ will depend at most on  the mean fields.  Therefore, we can expect
that a constant reflection coefficient will always remain a good approximation
close to equilibrium, namely  when $\Delta T/T \ll 1$ and $\Delta  n/n \ll 1$. 
In that case, the equations of the Sec.~\ref{sec:profiles} hold, regardless of
the average density of the system.

The  reader should  observe  that  $\alpha _\G^\Q$  and  $\alpha _\G^\J$  {\em
  differ},  while for  a system  without memory  effect (such  as  the Lorentz
model) one would expect equality  of these quantities.  This difference is not
an artifact of our simulations but a property of mechanical models of the type
we  consider.  Looking  at  Fig.~\ref{fig:geometry}, and  the  collision  rule
\eref{eq:coll-rule}, the reader will realize that fast particles have a higher
probability to  exit after only  1 collision with  the disc than  slower ones,
because  the angle  of  reflection tends  to be  smaller  for the  fast ones.  
Therefore the  memory effect  depends on the  individual energies and  not the
mean, and  this accounts  for the  difference above. We  have checked  that by
placing  non-rotating discs  between the  openings and  the turning  disc, the
effect is reduced.


\begin{figure}[!t]
\begin{center}
\includegraphics[scale=0.6]{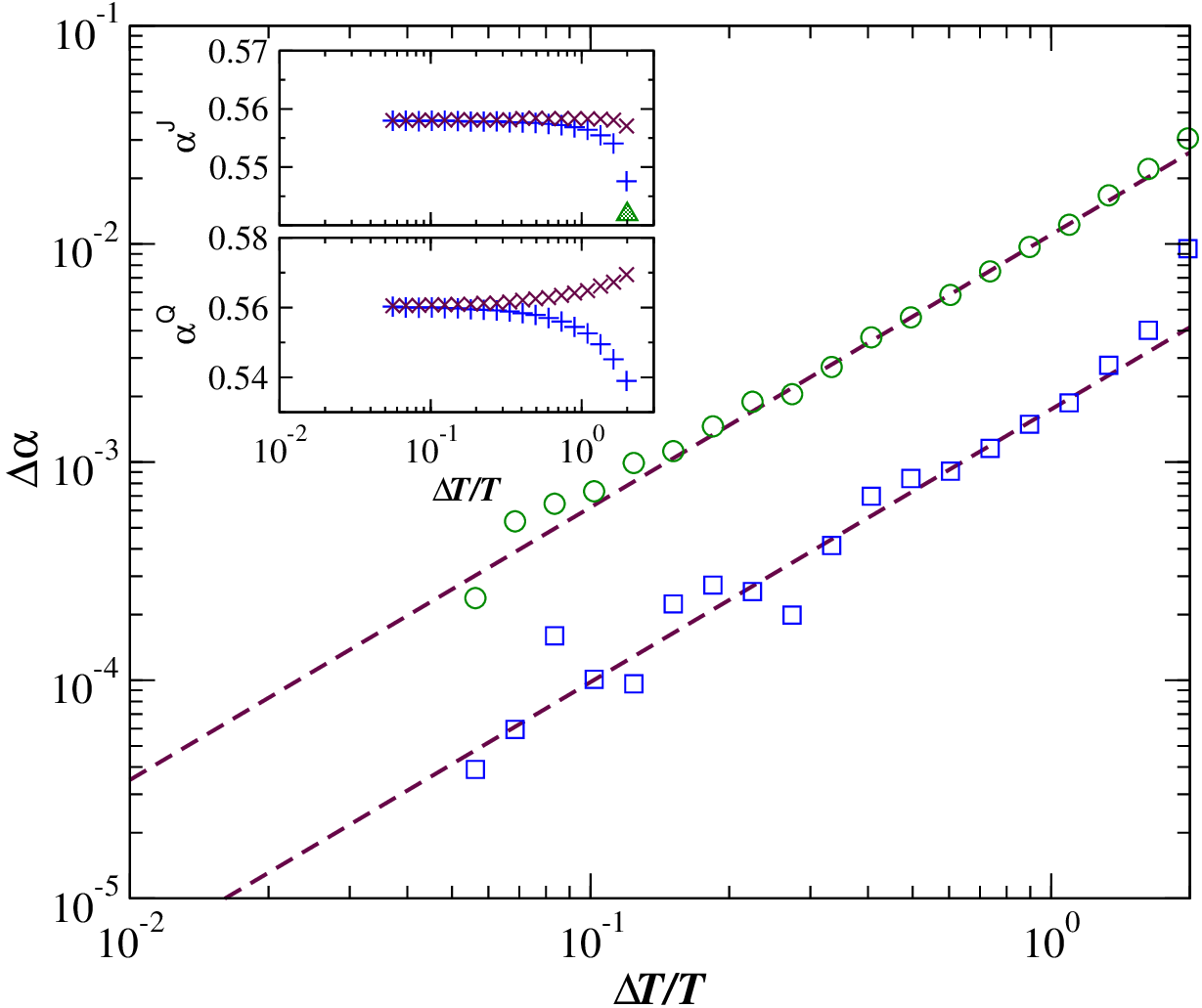}
\caption{Difference $\Delta\alpha=\alpha_\R-\alpha_\L$ as a function of
  $\Delta  T/T$  for $\AAJP$  (squares)  and  $\AAQP$  (circles). Each  symbol
  corresponds to  a simulation for the single  cell RDM with a  geometry as in
  Fig.~\ref{fig:th-profiles}  but  $|\gamma|=0.08$.   In all  experiments  the
  external gradients were fixed to $j_> =j_< =1$, $\Delta T=T_<-T_> = 100$ and
  $\HALF(T_>+T_<)=T$. The dashed  lines correspond to $\Delta\alpha\sim(\Delta
  T/T)^{5/2}$. In the  inset $\AAJP$ and $\AAQP$ at the  left (plus) and right
  (crosses) boundaries are shown.  Note that $\Delta\alpha\to0$ on approach to
  equilibrium,  while for  very large  gradients $\alpha  ^\J_\L$  reaches the
  Lorentz  gas limit  indicated by  the hashed  triangle (see  the text  for a
  discussion). \label{fig:alpha}}
\end{center}
\end{figure}

We now turn our attention to {\em nonequilibrium} effects, which are described
by the  functions $\epsilon  ^X_Y$ of Eq.~\eref{eq:alpha-cont}.   We performed
simulations for the single cell RDM  in which the external gradients were kept
fixed  but   the  mean  temperature   was  changed  ($T_>=T-\Delta   T/2$  and
$T_<=T+\Delta T/2$).  In the inset  of Fig.~\ref{fig:alpha}, we show  the left
and right reflection coefficients,  $\AA^\J_\L$ and $\AA^\J_\R$, as a function
of $\Delta T/T$. As the parameters approach equilibrium values, the difference
$|\AA^\J_\L-\AA^\J_\R|$  is  seen  to  decay  to  zero  (at  a  rate  $(\Delta
T/T)^{5/2}$).  The value that  $\AA^\J_\L$ and $\AA^\J_\R$ take in equilibrium
is the constant geometric asymmetry  (which for these parameters are $\AAJP_\G
\sim  0.558$ and  $\AAQP_\G \sim  0.560$).  At  the other  extreme,  since the
temperature  cannot   be  negative,  $\Delta  T/T  \le   2$.   The  difference
$\Delta\alpha$ is maximal at $\Delta T/T=2$.  Interestingly, $\AA^\J_\L(\Delta
T/T=2)$  corresponds  to  the  reflection  coefficient  of  the  Lorentz  gas,
\emph{i.e.}, to the limit in which the central disc does not rotate.

\section{Far from Equilibrium}
\label{sec:asymmetry}

In  Section~\ref{sec:profiles} we  have assumed  that the  system is  at local
equilibrium, and that the particle density varies in a range where the $\alpha
_\G$ are essentially constant. In this  section, we discuss the case when both
these  assumptions are  dropped.  This  means that  we take  into  account the
dependence of  the $\alpha$ on the density  and on the local  gradients of the
thermodynamical fields.

Far from equilibrium,  the distribution of the local  fields will be different
from cell  to cell  and thus,  the probabilities will  depend on  the position
along  the  chain  $\AA\equiv\AA(\xi)$.    We  recall  the  set  of  equations
Eq.~\eref{eq:alpha-cont} for $\alpha$ one of the functions $\alpha^X_Y$:
\begin{equa}
  \alpha (j_\L,j_\R,q_\L,q_\R) = \alpha _\G\left(\frac{j^{3/2}}{q^{1/2}}\right) +
 \epsilon \left (\frac{j^{3/2}}{q^{1/2}},\frac{j_\R-j_\L}{j_\R+j_\L},\frac
{q_\R-q_\L}{q_\R+q_\L}\right ) \ .
\end{equa}
It is  understood that $\alpha_\G$  and $\epsilon$ denote functions  which are
different  for  the  various  $\alpha^X_Y$,  but  have  the  following  common
properties: By definition, $\epsilon$ must vanish at equilibrium
\begin{equ}
\epsilon^X_Y(x,0,0) \ = \ 0 \ .
\end{equ}
Furthermore, the left-right symmetry of the cell implies:
\begin{equ} \label{eq:LR-sym}
\epsilon^X_\L(x,y,z)=
\epsilon^X_\R(x,-y,-z)
 \ .
\end{equ}

We  next  discuss specific  properties  of $\epsilon  $  as  they appear  from
numerical simulations for the RDM.  It  turns out that the $\epsilon ^X_Y$ not
only vanish  on the submanifold $(x,0,0)$,  but, to a  very good approximation
they  vanish for those  points where  the temperatures  $T_\L$ and  $T_\R$ are
equal.   Since  the temperatures  are  functions  of  the injection  rates  of
particles  and   energy  this  means  that  $\epsilon^X_Y$   vanishes  on  the
submanifold of $\{j_\L,j_\R,q_\L,q_\R\}$ where  $T_\L=T_\R$.  While we have no
proof  of  this  observation, it  can  be  understood  by observing  that  for
$T_\L=T_\R$  the tank  has  the same  temperature  as all  the particles,  and
therefore we are  in presence of purely geometric  phenomena which are already
captured by the function $\alpha _\G$ alone.

We expand $\epsilon^X_Y$ to first order, and in view of the above information,
a very good approximation is given by assuming that $\epsilon$ has the form:
\begin{equ}\label{eq:first-order}
  \epsilon \,=\, \epsilon_0 \frac{T_\R-T_\L}{T_\R+T_\L}+\epsilon_1\frac{T_\R-T_\L}{T_\R+T_\L}\cdot\frac{j_\R-j_\L}{j_\R+j_\L}~,
\end{equ}
where $\epsilon  _0$ and $\epsilon _1$  are constants depending  on the choice
$X$, $Y$.
\begin{figure}[!t]
\begin{center}
 \includegraphics[scale=0.6]{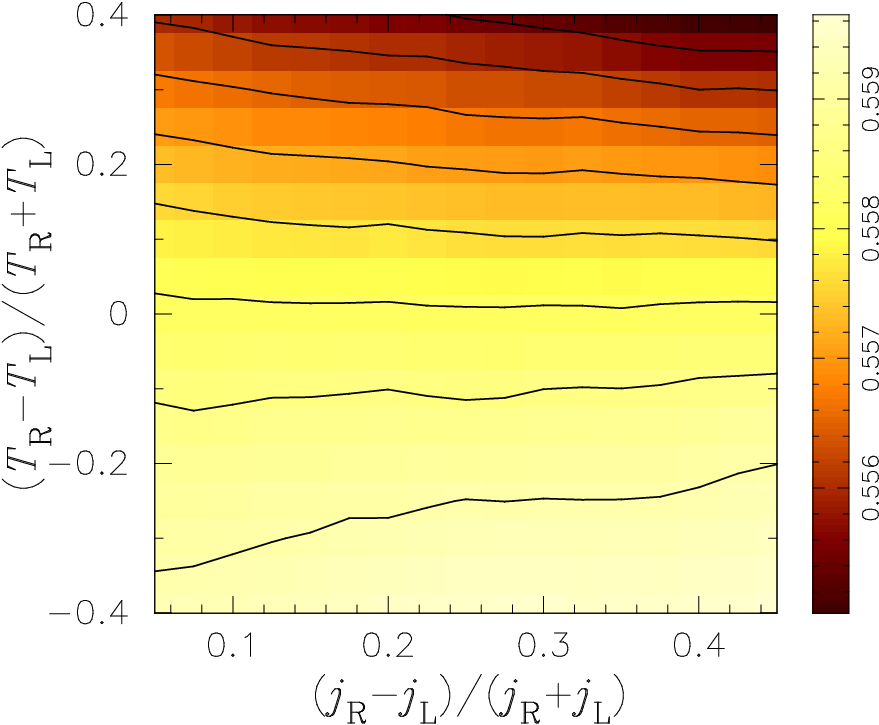} 
 \caption{ Contour density plot of $\alpha^\J_\L$ as a function of the local
   gradients in  $j$ and $T$. The  obtained values correspond  to the one-cell
   RDM  with an  opening  of $\gamma=0.08$.  Note  that the  contour line  for
   $T_\L=T_\R$              is              horizontal,             justifying
   Eq.~\eref{eq:first-order}.\label{fig:rhoL}}
\end{center}
\end{figure}
We  have numerically  corroborated Eq.~\eref{eq:first-order}  for the  RDM. In
Fig.~\ref{fig:rhoL} a  contour density plot  for $\alpha^\J_\L$ is shown  as a
function of  the local  gradients.  The linearity  of the contours  shows that
$\alpha^\J_\L$ is a  linear function of the local gradient  in $j$ whose slope
and intercept depends on the  local gradient of temperature. The same behavior
is found for the other coefficients.

The reader  should note that for  fixed external parameters (for  the heat and
particle reservoirs) and  for long chains, the local  gradients per cell scale
like $1/N$  and thus, one  expects the $\epsilon$  to be both  proportional to
$1/N$ and to the force fields. The  question then is whether the effect of the
asymmetry will disappear in the infinite volume limit or not.
 
We next  show how,  in principle,  the above approximations  lead to  a closed
system of  equations for the profiles,  and we then check,  for one particular
case  that the  equations  indeed  describe the  profiles  found in  numerical
experiments,    and    account    correctly    for    the    deviation    from
\cite{eckmann-young}.

Since the  local gradients are naturally  expressed in terms  of the injection
rates we use the identification  between injection and ejection rates to write
Eq.~\eref{eq:balance} as
\begin{equa} \label{eq:balance-inj}
j_{\R,k-1} = \alpha^\J_{\L,k}~j_{\L,k} + (1-\alpha^\J_{\R,k})~j_{\R,k}~,\\
j_{\L,k+1} = (1-\alpha^\J_{\L,k})~j_{\L,k} + \alpha^\J_{\R,k}~j_{\R,k}~,\\
\end{equa}
with $k=1,\dots,\N$ and boundary conditions $\jL{1}=j_>$ and $\jR{N}=j_<$.

Using the definition of the particle current $\varphi^\J=j_{\L,k+1}-j_{\R,k}$,
one    can    eliminate   the    $j_\R$    (or    the    $j_\L$),   and    the
Eqs.~\eref{eq:balance-inj}  in  which  case  the  system reduces  to  only  on
equation:
\begin{equ} \label{eq:balance-L}
(1-\alpha^\J_{\L,k})~j_{\L,k}-(1-\alpha^\J_{\R,k})~j_{\L,k+1}+\alpha^\J_{\R,k}~\varphi^\J=0 \ .
\end{equ}
Analogous expressions are obtained for the energy injection rates.

It  is here  that  the  non-gradient nature  of  our models  is  visible in  a
nutshell.  For  \eref{eq:balance-L} to be of  gradient type one  needs to have
$\alpha^\J    _{\L,k}    =\alpha^\J_{\R,k}$.     As    we   have    seen    in
Fig.~\ref{fig:alpha}, this is, in general, not the case.

\subsection{Infinite volume limit}

We take  the continuum  limit of Eq.~\eref{eq:balance-L},  with cells  of size
$1/N$ so  that the rescaled variable  $\xi=k/(N+1)$ is in the  domain $[0,1]$. 
Note that the currents $\varphi^\J$ and $\varphi^\Q$ scale (for fixed external
forces) like ${1}/{N}$.

In order to  close the balance equation for $j_\L$  one substitutes the ansatz
\eref{eq:first-order} into Eq.~\eref{eq:balance-L}. This leads to an involved,
but in  principle straightforward system of  nonlinear differential equations,
which we  do not write down.  However,  we will deal with  the special simpler
case in which $q(\xi)$ is approximately constant along the chain.

When  $q$  is  constant, one  finds  that  $T\propto  1/j$ and  therefore  the
assumption  \eref{eq:first-order} can  be reformulated  for  the corresponding
$\epsilon_\L$ and $\epsilon_\R$ to first order in $1/N$ as
\begin{equa}[4] \label{eq:asymm-hypo} 
&\eL{k}^\J& \ = \ & \A^\J\, \frac{\jR{k} - \jL{k}}{\jR{k} + \jL{k}}~,\quad
&\eR{k}^\J& \ = \ & -\A^\J\,\frac{\jR{k} - \jL{k}}{\jR{k} + \jL{k}} \ ,\\
\end{equa}
where $\A^\J=-\epsilon _0$ are constants determined by the boundary conditions
of  \eref{eq:balance-L}. The  minus sign  in the  equation for  $\eR{k}$  is a
direct consequence of \eref{eq:LR-sym}.

\begin{figure}[!t]
\begin{center}
 \includegraphics[scale=0.5]{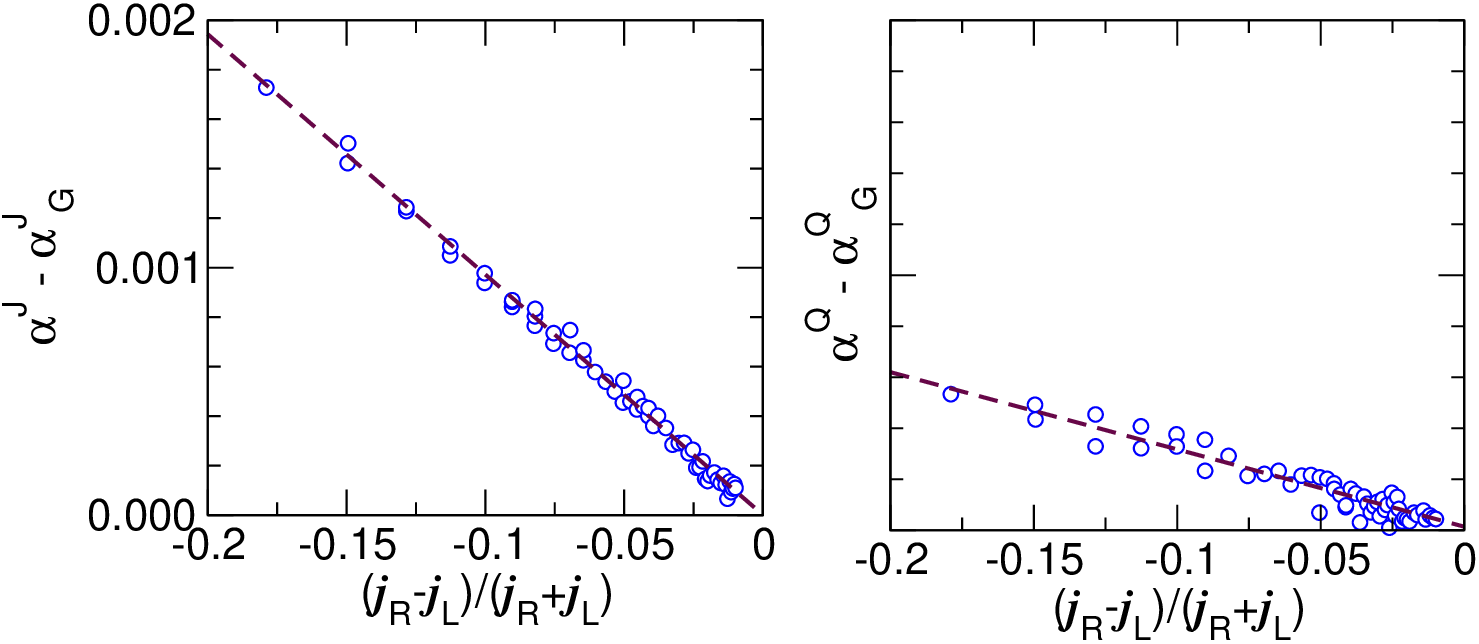} 
 \caption{ Dependence of the left (squares) and right (circles) reflection
   probabilities $\AA^\J-\alpha  _\G^\J$ and  $\AA^\Q_\G$ on the  gradients of
   the local  fields for  a chain  of $20$ RDM  cells with  $\gamma=0.08$. The
   external gradients  were fixed  to $j_>=1$, $j_<=6$,  and $q_>=  q_<=1500$. 
   The  values  for  the  $\alpha  _\G$  have been  taken  from  the  data  of
   Fig.~\ref{fig:n-dep}.
 \label{fig:grad-dep}}
\end{center}
\end{figure}

We  have  numerically corroborated  the  hypothesis \eref{eq:asymm-hypo}.   In
Fig.~\ref{fig:grad-dep}   we   show   the   dependence  of   the   asymmetries
$\epsilon^\J$ and $\epsilon^\Q$ on the  gradients of the local injection rates
for a chain of $20$ RDM cells.  The asymmetries are seen to depend linearly on
the local gradient  of $j$.  Farther from equilibrium  (not shown), deviations
will appear.

Using the  particle current we  express the asymmetry  \eref{eq:asymm-hypo} in
terms of the $j_\L$ alone as
\begin{equa}[2] \label{eq:asymm-J} 
&\eL{k}^\J& \ = \ & \quad \A^\J \frac{\jL{k+1}-\jL{k}-\varphi^\J}{\jL{k+1}+\jL{k}-\varphi^\J}~,\\
&\eR{k}^\J& \ = \ & -\A^\J \frac{\jL{k+1}-\jL{k}-\varphi^\J}{\jL{k+1}+\jL{k}-\varphi^\J}\ .
\end{equa}

Finally,  inserting  \eref{eq:asymm-J} into  \eref{eq:balance-L}  we obtain  a
closed  equation for  $j_\L$ with  only one  system dependent  free parameter,
$\A^\J$.  In the scaling limit  one finds the following nonlinear differential
equation for the injection rates:
\begin{equa} \label{eq:continuous}
j'_\L(\xi) &= - \frac{ \alpha_\G^\J\bigl(n(\xi)\bigr)-\A^\J}{1
-\alpha_\G^\J\bigl(n(\xi)\bigr)+\A^\J}\,(N\varphi^\J)+\OO\left(\frac{1}{N}\right)~.\\
\end{equa}
The same equations hold for  the ``$\R$'' versions but with different boundary
conditions.

Note that $\varphi^\J$ is  the current of a chain of length  $N$, and hence is
asymptotically  equal to  $\Phi^\J/N$.  Using  \eref{eq:continuous}, $\Phi^\J$
can be determined by the boundary conditions as
\begin{equ}
\Phi^\J \,=\,\frac{j_>-j_<}{\int_0^1 W^\J(\xi)d\xi}~,
\end{equ}
where    $W^\J(\xi)$     is    the    coefficient     of    $\varphi^\J$    in
Eq.~\eref{eq:continuous}.

\begin{figure}[!t]
\begin{center}
 \includegraphics[scale=0.53]{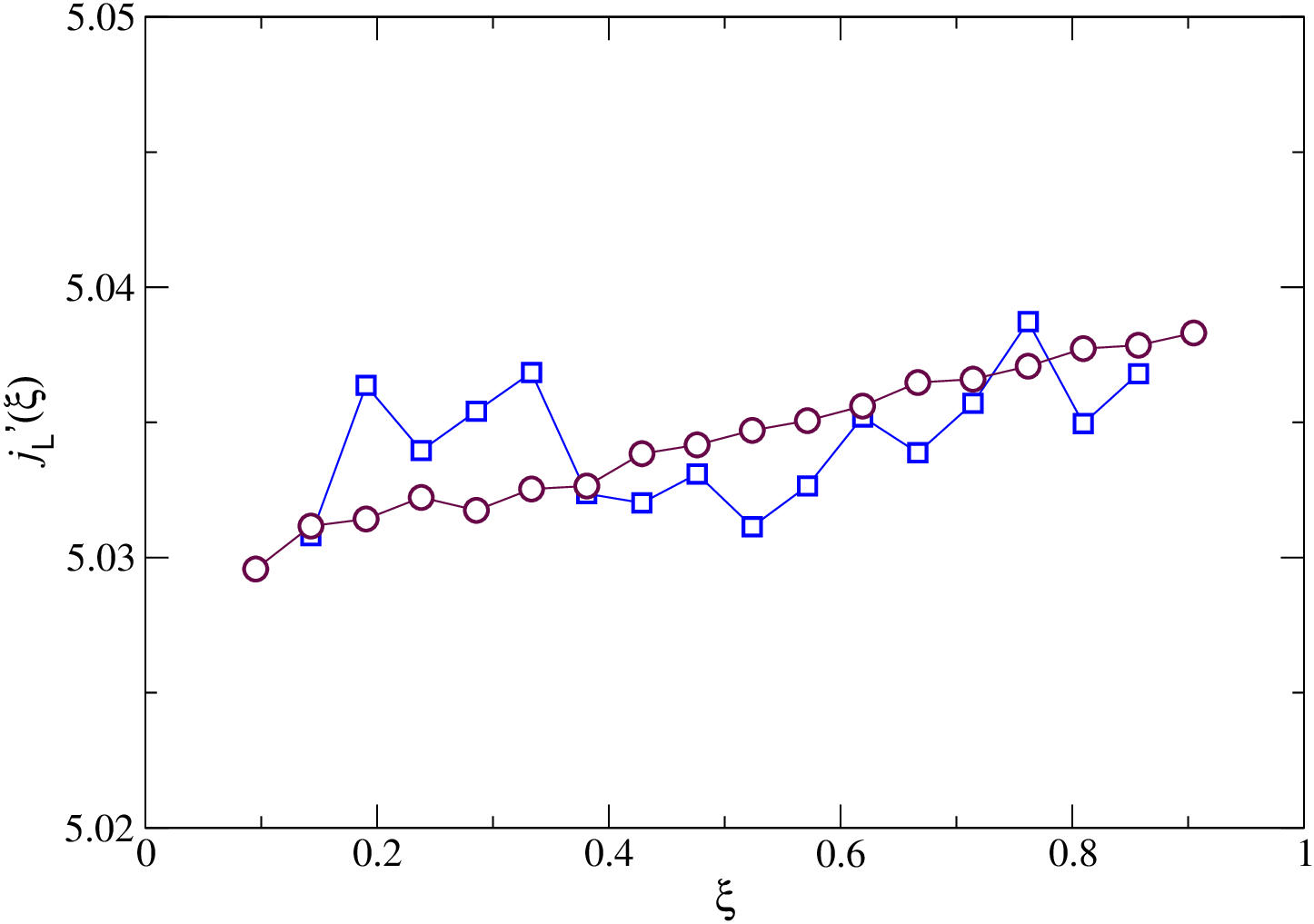} 
 \caption{ Numerical verification of Eq.~\eref{eq:continuous}, for $j_\L'$, for
   systems   of  size   $N=20$.   The  external   parameters   are  those   of
   Fig.~\ref{fig:grad-dep}.    The  circles   correspond   to  the   r.h.s.~of
   \eref{eq:continuous} as explained in the text, while the squares correspond
   to $(j_{\L,k+1}-j_{\L,k})(N+1)$--- measured  at $\xi=k/(N+1)$---which is an
   approximation to $j'_\L(\xi)$.\label{fig:infty-limit}}
\end{center}
\end{figure}

As discussed  in Sec.~\ref{sec:bias} (see  \emph{e.g.}, Fig.~\ref{fig:n-dep}),
we have  found that at sufficiently  low or high densities,  the $\AA$ depends
little on the local fields, taking practically a constant value. Therefore, if
the  external gradient imposed  by the  injection of  the reservoirs  into the
system is such  that the density in all  the cells of the chain  is either low
($n\lesssim0.3$) or  high ($n\gtrsim10$)  then the $\alpha_\G(n(\xi))$  can be
taken as constant along the chain.  In this situation Eq.~\eref{eq:continuous}
gives    a    linear    profile    for    the    injection    rate,    $\d_\xi
j_\L(\xi)=\mathrm{const}$.  In particular, in  the high density regime, linear
profiles  will be  obtained even  when the  gradients are  very large  and the
system is far from equilibrium.

For  intermediate   densities,  it  follows   from  \eref{eq:alpha-cont}  that
$\alpha_\G$ is a function of
\begin{equ}
\sqrt{\frac{(j_\L+j_\R)^3}{q_\L+q_\R}} \ ,
\end{equ}
which to lowest order in ${1}/{N}$ is
\begin{equ} \label{eq:bias-dev}
n(\xi)\propto \sqrt{\frac{j_\L^3(\xi)}{q_\L(\xi)}} \ .
\end{equ}

Therefore,  in the  intermediate  density  regime, where  we  have found  that
$\alpha_\G(n)$ is a linear  function of $n$, Eq.~\eref{eq:continuous} leads to
a system of two explicit coupled differential equations for $j$ and $q$, which
can be solved.

The  consistency  of our  approximations  was checked  for  the  RDM model  by
combining   the   measurements   of    all   the   quantities   appearing   in
Eq.~\eref{eq:continuous}.   The  results  for  the discrete  version  of  this
equation  are  summarized  in  Fig.~\ref{fig:infty-limit},  which  shows  good
agreement  with  the  theory.    These  measurements  also  show,  that  while
correlations are to be expected for high gradients, \cite{spohn,nicolis}, they
do not seem to affect the validity of our approximations.

The  external  parameters used  are  those  of Fig.~\ref{fig:grad-dep},  which
guarantee  that all  densities along  the chain  lie in  the domain  of linear
dependence  shown in Fig.~\ref{fig:n-dep},  \emph{i.e.}, between  0.3 and  10. 
The  data   in  Fig.~\ref{fig:infty-limit}  were  obtained   as  follows:  The
derivatives  $j'_\L$ were  approximated  with centered  differences along  the
profiles.    The   $\alpha   _\G$   was   read   off   the   linear   fit   of
Fig.~\ref{fig:n-dep}.    The  flux   $\varphi$  is   directly  read   off  the
simulations,  while the  $\A$  coefficients  are obtained  from  the slope  in
Fig.~\ref{fig:grad-dep}.

\section{Conclusions}
\label{sec:conclusions}

Guided  by   a  stochastic   description,  we  developed   a  phenomenological
description of heat and matter transport for a family of Hamiltonian models of
many interacting particles.  The description is based on a careful analysis of
the way in which particles and energy enter and leave the individual cells.

A  stochastic model  for  a  Hamiltonian system  must  include an  appropriate
treatment of the memory effects  that arise due to the deterministic character
of the dynamics.   This approach would be cumbersome.   Instead, we followed a
different approach that,  inspired by an approximated stochastic  model,  is
based on  the phenomenology  of reflection probabilities  of particles  and of
energy in terms of the local fields and the local gradients.

We  have expressed a  set of  balance equations,  accounting for  particle and
energy  conservation  in   the  steady  state  in  terms   of  the  reflection
probabilities.    After   specifying   a   phenomenological  law   for   these
probabilities we obtain closed expressions  for the local fields and show that
they  capture the  essential features  of the  microscopic  dynamics including
memory effects inherent to any Hamiltonian deterministic system.

A useful observation was to  identify two contributions to the memory effects:
one of geometrical and another  of dynamical origin.  Close to equilibrium the
geometric component  dominates and the corresponding  correction preserves the
``gradient''  character of  the system.   The theory  remarkably  predicts the
transport properties of our class  of Hamiltonian models, namely the fluxes of
heat and matter and the steady state energy and density profiles.

Far  from equilibrium,  the dynamical  effects  strongly depend  on the  local
gradients  of   the  thermodynamical  fields  and  thus,   on  the  particular
interaction between particles  and energy tanks.  Finally, we  have shown that
in the  continuum limit the system is  no longer of ``gradient''  type and the
energy and particle  currents are not proportional to  the external gradients. 
We   have   obtained  the   lowest-order   deviation   from   the  theory   in
\cite{eckmann-young}.  However, we have found that the gradient type condition
is restored in the limit of very high or very low densities.

Given that the dynamical memory effects depend on the particular nature of the
interaction,  our  explicit  solution  (\emph{e.g.}, the  particular  case  of
\eref{eq:continuous}), cannot  be applied to a general  context.  However, the
program  outlined and  the  variables used  in  our derivation  are valid  and
appropriate for the family of models that we have considered.

\subsection*{Acknowledgments}
We  thank  G. Jona-Lasinio  for  enlightening  discussions  and the  anonymous
referees for  useful questions  and remarks concerning  an earlier  version of
this paper.  This work was partially supported by the Fonds National Suisse.

\end{document}